\documentstyle[epsfig]{l-aa}
\title{Inner accretion disk disappearance during a radio flare in GRS 1915+105}
\author{M. Feroci\inst{1} 
\and G. Matt\inst{2}
\and G. Pooley\inst{3} 
\and E. Costa\inst{1}
\and M. Tavani\inst{4,5}
\and T. Belloni\inst{6,7}
}

\begin{document}
\offprints{feroci@ias.rm.cnr.it}
\date{Received May 15, 1999; accepted September 28, 1999}
\thesaurus{011 (13.07.1; 13.25.1)}

\institute{{Istituto di Astrofisica Spaziale, CNR, 
Via Fosso del Cavaliere, I-00133 Roma, Italy}
\and
{Dipartimento di Fisica, Universita' `Roma Tre', 
Via della Vasca Navale 84, I-00146 Roma, Italy}
\and
{Mullard Radio Astronomy Observatory, Cavendish Laboratory, Madingley Road,
Cambridge CB3 0HE, UK}
\and
{Istituto Fisica Cosmica e Tecnologie Relative, CNR, Via Bassini 15,
I-20133 Milano, Italy}
\and
{Columbia Astrophysics Laboratory, New York, NY, USA}
\and
{Astronomical Institute "A. Pannekoek" and Center for High-Energy
Astrophysics, Kruislaan 403, 109855 Amsterdam, The Netherlands}
\and
{Osservatorio Astronomico di Brera, Via Bianchi 46, I-23807 Merate, Italy}
}

\maketitle

\begin{abstract}

Simultaneous X--ray (BeppoSAX, 0.1--300~keV)
and radio (Ryle Telescope, 15~GHz) observations of the
superluminal galactic source GRS 1915+105 in high flux state
were carried out starting October 1996. 
The source was in a phase of large X--ray intensity and spectral variability.
Among the simultaneous observations, an isolated radio flare lasting
$\sim$2000~s was detected accompanied by X--ray outbursting activity.
Here we present the simultaneous radio/X--ray light curves 
of the flare event and the spectral and timing analysis of 
the relevant BeppoSAX data.
The broad band (0.1-200 keV) energy spectrum of the source can be 
described in terms of the `standard' spectral model for black hole 
candidates. 
We obtain evidence for a temporary disappearance and subsequent 
restoring of the inner portion of the accretion disk during the flare
from the spectral analysis of the X--ray data. 
We find a  variable power law photon index positively
correlated to the variability of the exponential cut-off energy.
In addition, frequency and intensity variations of a $\sim$4-5~Hz 
quasi-periodic oscillation (QPO) are detected during the same time interval.
The centroid frequency of the QPO is found to be correlated to the
spectral power law photon index, and to the X--ray flux from  
the disk and the high energy flux from the power law.
These observational results are discussed 
in the framework of the disk-instability model proposed for this source. 


\end{abstract}

\section{Introduction}

GRS 1915+105 was discovered in X--rays in 1992 by the WATCH
experiment onboard GRANAT (\cite{Castro}). It was then
observed in the radio band, where plasma ejection activity with 
apparent superluminal motion was observed (\cite{mirabel94}).
This jet source, sometime referred to as a {\it microquasar} (due
to a phenomenology analogous to a quasar but scaled down in mass
by several orders of magnitude),
is located close to the galactic plane, at a distance of $\sim$12 kpc 
as inferred from kinematic and HI measurements (\cite{mirabel94,fender99}). 

In X--rays GRS 1915+105 shows a wide range of variability on any sampled time
scale. Quasi-periodic oscillations have been observed with the PCA
experiment onboard the {\it Rossi X-ray Timing Explorer} (\cite{greiner97})
at frequencies varying from a few mHz to 67 Hz (\cite{morgan97}).
The latter frequency, unlike all the others, 
does not change with time, and has therefore been
suggested to be related with the innermost stable orbit of the accretion
disk around the black hole (\cite{morgan97}). 

The X--ray spectrum of the source has been successfully described 
in terms of emission from an optically thick accretion disk 
(Belloni et~al. 1997a). The disk emission is modelled 
as a multi-temperature blackbody of innermost
temperature $\sim$1-2~keV. A population of energetic electrons  
can Compton up-scatter soft photons from the disk, thus originating 
the power law component observed in the spectrum.  
A low energy absorption is detected in X--ray observations,
corresponding to a neutral hydrogen column of
$\sim$5x10$^{22}$ atoms cm$^{-2}$ (\cite{greiner97}); the
Galactic neutral hydrogen column to the source direction
is $\sim$2x10$^{22}$ atoms cm$^{-2}$ (\cite{nh}).
The extreme variability of the source (both in flux and energy spectrum) 
has been interpreted (\cite{belloni97a}) in terms of a cycle consisting of
the abrupt emptying of the inner portion of the accretion 
disk followed by a slower refilling (\cite{belloni97b}).

Radio maps have shown that during its active phases the source often 
exhibits events of plasma ejections, observed as extended radio flux
emission. 
In some cases the outflowing plasma propagates through the interstellar
medium for tenths of a parsec, with an intrinsic velocity of
0.92-0.98~c and apparent superluminal motion (\cite{mirabel99,fender99}).
Using simultaneous observations at different wavelengths, 
several authors 
(\cite{pooley97,fender97,fender98,eikenberry98a,mirabel98,eikenberry98b,fender99})
were able to correlate the X--ray variability with the emission
at radio and infrared frequencies. In particular, they
showed that the observed multi-frequency behaviour is consistent
with a scenario in which instabilities in the inner portion of the
accretion disk, detected as oscillations of the X--ray flux, 
cause a plasma jet ejection
that adiabatically expands away from the source. 
The radio and infrared radiation detected from the jets is consistent with
a synchrotron emission mechanism.

\begin{figure*}
\centerline{\epsfig{file=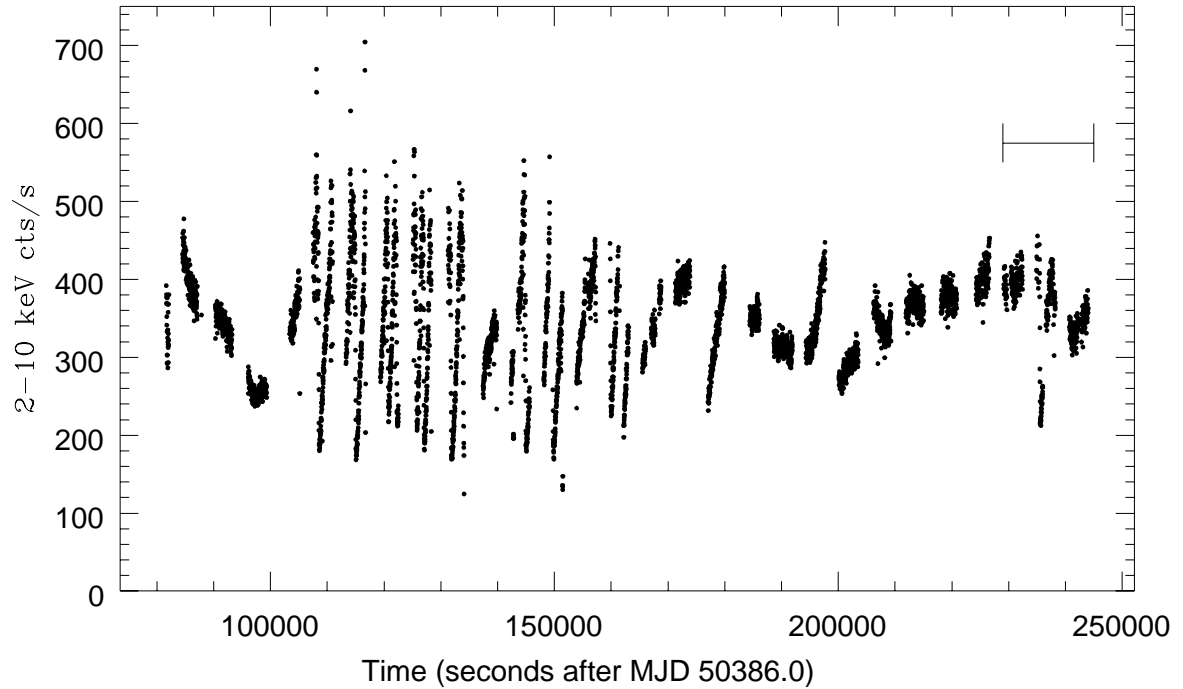,width=17cm}} 
\vspace{-6cm}
\caption[]{BeppoSAX MECS (2--10~keV, bin size 16 s) light curve of the long
observation of GRS 1915+105, showing the source behaviour before
the occurrence of the outburst shown in figure \ref{LC_flare}. 
The time interval shown in Figure \ref{LC_flare} 
is indicated here with an horizontal line.
Data gaps are due to Earth occultation and to 
passages of BeppoSAX close to the South Atlantic Anomaly.}
\label{LC_total}
\end{figure*}

\section{Observations}

A Target of Opportunity observational campaign (\cite{feroci98,matt98}) 
with the Narrow Field Instruments
(NFI: LECS, MECS and PDS, \cite{boella97}) onboard BeppoSAX 
started after its Wide Field Cameras (2--26~keV, \cite{jager97})     
detected intense X--ray activity from this source. 
After two observations of about 10 ks, the source was observed
continuously with the NFI for more than one day (Fig.~\ref{LC_total}), 
and then monitored
again for two more weeks with a series of short observations. 
Along with the BeppoSAX observations, radio monitoring 
of GRS 1915+105 was performed at the Ryle Telescope at 15 GHz
(\cite{pooley97}), resulting in a sparse set of short X--ray/radio 
simultaneous observations. 

In this paper we discuss one of such observations, 
during which a radio flare was detected
simultaneously with an isolated X--ray outbursting activity of the source. 
The X-ray spectral and timing characteristics are of 
particular interest and, as we will show, can be interpreted in terms
of the sudden disappearance of the inner portion of the accretion disk.
The analysis of the remaining observations will be presented in a 
forthcoming paper.

\subsection{The flare event}

\begin{figure}
\epsfig{file=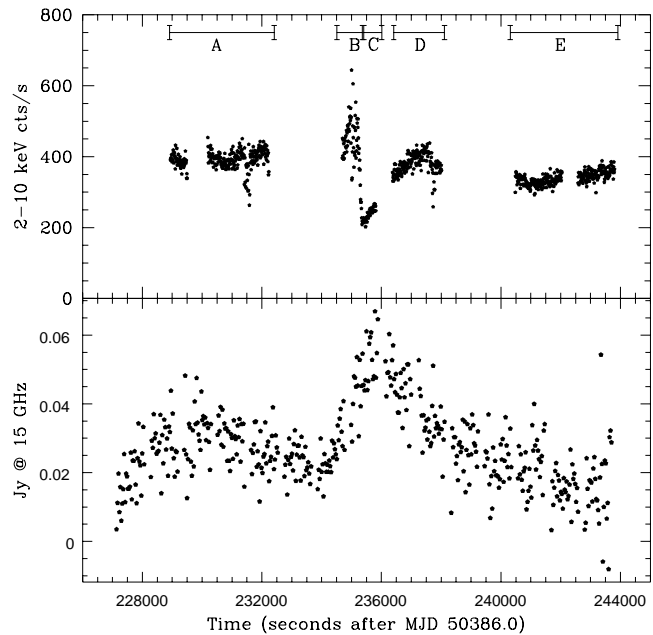,width=9cm}
\caption[]{BeppoSAX MECS (2--10~keV, bin size 10~s) 
and Ryle Telescope (15~GHz, 32~s bin size) 
light curve of the flare from GRS 1915+105. The data gaps in the X--ray light
curve are due to the satellite passage near the South Atlantic Anomaly
and to Earth occultations.}
\label{LC_flare}
\end{figure}

\begin{figure}
\epsfig{file=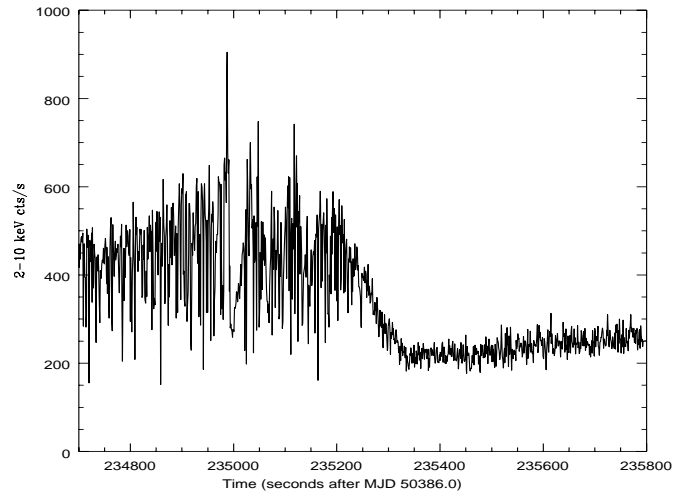,width=9cm,height=7cm}
\caption[]{Zoom-in (bin size 1 s) of the MECS light curve 
of the outburst reported in 
Fig.~\ref{LC_flare}, showing the complex timing behaviour of the source 
during the time intervals B and C. For this figure it looks very likely 
that the high frequency oscillations started earlier than the beginning 
of interval B, during the Earth occultation}
\label{LC_zoom}
\end{figure}

On 1996 November 1.628 UT, GRS 1915+105 exhibited the X--ray outburst shown 
in Fig.~\ref{LC_flare} (top panel) and \ref{LC_zoom}.
The source was in a high-flux ($\sim2.3\times 10^{-8}$ erg cm$^{-2}$ s$^{-1}$ 
in 2--10~keV), relatively quiet state when the outburst occurred;
during the outburst large amplitude 
oscillations were observed (Fig.~\ref{LC_zoom}). After that, 
the source went into a dip and finally recovered a state 
similar to that before the outburst. On a longer timescale, the event occurred 
after a period of relatively quiet X-ray behaviour (at high flux level) 
following an impressive flaring state lasting 
almost one day (Fig.~\ref{LC_total}, \cite{feroci98}). 
This temporal behaviour may be seen in context of the scenario described in 
Belloni et~al. (1997a). 
At the time of the X--ray outburst the inner portion of the disk
undergoes an instability that, in few hundred seconds, 
leads to its disruption, causing a decrease in the emitted X--ray flux. 
The instability results in ejection of plasma out of the inner disk. 
Once the instability is damped, the inner accretion disk is replenished 
and the X--ray emission restored.
 
The 15~GHz radio emission during the same time period is shown in the 
bottom panel of Fig.~\ref{LC_flare}. 
At the time of the X--ray outburst (in a period unfortunately not fully 
covered by the X--ray observation due to the Earth occultation), 
the radio emission 
shows a smooth, relatively fast rise followed by a slower decay. 
The radio flux increase is almost linear with time,
reaching a peak flux of $\sim$ 60~mJy in correspondence with the X--ray dip. 
The decay from this peak follows a quasi-exponential law down to a flux
of $\sim$ 20 mJy, when the radio observation stops.  
The time profiles in the two bands are rather different, being more
complex in X-rays than in the radio. 

In the correlated X--ray and radio events studied so far
(\cite{pooley97,mirabel98,eikenberry98b}) the source exhibited a
series of X--ray flares and dips. The corresponding
radio and infrared flares have been interpreted 
(e.g., \cite{mirabel99} and references therein)
in terms of emission from
plasmoids ejected at the time of the preceding X--ray dips, with the
relative X-ray/infrared/radio delay due to the optical thinning during
adiabatic expansion of the plasma bubbles.
However, we note that in that observations the relative phase 
between the X--ray and radio/IR
light curves is not always the same. In some cases (see Fig.~2 of
Mirabel et al. 1998) the radio peak is reached during the X--ray outburst,
but in several other cases (see Fig.~5 of Pooley \& Fender 1997, 
Fig.~1 of Mirabel et al. 1998 and Fig.~2 
of Eikenberry et al. 1998, if one extrapolates that the radio peaks after
the IR) the radio emission peaks during the X-ray dip, just like we observe.
Of course, these differences can be attributed to casual superpositions
of uncorrelated events, due to the intense activity of the source. 
In this scenario, it is possible that the correlation between 
X--rays and radio emission that we observed (i.e. the radio peak 
at the time of the X--ray dip) is also fortuitous, and the radio 
event is actually associated with the dip of an unseen X--ray flare 
occurred during the data gap due to Earth occultation 
(i.e. between intervals A and B in Fig.~\ref{LC_flare}). 
However, we note that in this case one would 
naturally expect also a radio event following the observed X--ray event. 
This is not seen, unless this putative radio flare is blended with 
the decay of the main radio event. 
The latter hypothesis would in turn likely require that
the unseen X--ray outburst is much larger than the 
observed one (if a relation exists between the X--ray flare 
intensity and the radio emission intensity). 

Alternatively, if we use only the information directly derived from our
observations and assume that the radio flare is actually associated 
to the observed X--ray flare, 
we may suppose that we are dealing with a somewhat 
different event with respect to those interpreted in the 
literature, or with the same phenomenon on a smaller scale
(and therefore a shorter time scale). 
This may not be too much surprising also looking at the context 
in which the X--ray flare occurred (Fig.~\ref{LC_total}): 
it appears to be rather isolated in a period of relative stillness 
of the X--ray emission, while the events usually discussed
in the literature occurred in periods of high X--ray variability, 
within long and continuous sequences of flares and dips 
individually similar to the one we show here 
(i.e., a period like the first half of our long observation,
Fig.~\ref{LC_total}). 
In fact, it is known that some radio flares from this and similar 
sources are not related to substained plasma emission forming a 
steady jet, being instead exhausted at an early stage 
(`aborted events', that is, small versions of the big ejections with 
spectacular jets; 
R. Hjellming, private communication). 
Suppose we are dealing with one of these `aborted jets',
during the high frequency oscillations (interval B and earlier) 
relatively small plasmoids are ejected out of the disk, 
rapidly becoming optically thin and being therefore observable 
at radio frequencies. 
In this scenario, the observed radio peak would simply be the convolution 
of individual plasmoid's emission ejected in a sequence which ceased at
the time of the
X--ray dip (when the inner accretion disk disappears, see below), 
causing the subsequent decrease in the radio flux. 

We assume this scenario as our working hypothesis.
In the next section we will test the inner disk disappearance hypothesis 
using a spectral modeling of the 0.1--200~keV X--ray emission observed 
with BeppoSAX during the event. Then, in the following section we search 
for timing signatures in the X--ray data.

\section{Broad band spectral analysis}

In order to infer the source behaviour from the BeppoSAX spectral
data we divided the observation in five time intervals (A, B, C, D, E)
as shown in Fig.~\ref{LC_flare}. 
The aim of this time selection was to divide the event on the basis of
the source state. 
We note, however, that the chosen time intervals
could include significant spectral variation of the source,
as suggested by Fig.~\ref{LC_zoom} for interval B, and
the results that we derive could represent `average states'
of the source.   
 
The available spectral data from the BeppoSAX observation are from the
LECS instrument (0.1--4~keV) for portions of the A, D and E intervals; 
from the MECS (1.8--10~keV) and from the PDS (15--200~keV) for all
the five intervals.
In the spectral analysis with these three instruments we used a
normalization factor for the intercalibration with the MECS
(\cite{cookbook}): 0.84 (fixed) for the PDS, 
free parameter for the LECS (usually between 0.9 and 1.0).  
The LECS data have been mainly used to determine the column density of the 
absorbing matter, which has been found to be constant in the three 
available time intervals at the value of $\sim$5.6$\times10^{22}$ cm$^{-2}$.
Since LECS data are available only for portions of only 3 of the
5 selected intervals, in order to avoid systematic differences in the data 
analysis between spectra with and without the LECS we 
fitted only MECS and PDS data, with the neutral hydrogen column density 
fixed to the value derived from the LECS.

For the spectral fitting (performed with the 
XSPEC v. 10 package, \cite{arnaud96}) we used
a model including the following components: 
a multi-temperature disk blackbody, a power
law with a high energy exponential cut-off,
an iron emission line and a Compton reflection
continuum from cold matter (assuming an inclination angle of 70$^{\circ}$).
This model was typically used for black 
hole candidates (\cite{mitsuda84}),
and it is the same used in \cite{belloni97a} to fit RXTE/PCA data,
except for the high energy cut-off and the Compton reflection that 
we added due to the wider (0.1--300~keV) BeppoSAX spectral coverage. 
The use of the Compton reflection component (with the high energy 
cut-off included in the relevant {\it XSPEC} model) 
is generally preferred
to a simple cut-off power law when fitting the BeppoSAX wide 
band energy spectra 
(not only those presented in this paper) of this source.
Actually, it is also used to fit RXTE data when also the high energy 
instrument, HEXTE, is used (\cite{greiner98}).
However, this component is not always well constrained. As an example, 
in case of interval E if we change the Compton reflection to 
a cut-off power law, we pass from $\chi^{2}$=149.1 (89 d.o.f.) to
$\chi^{2}$=161.7 (90 d.o.f.), and 
an F-test attributes a probability of 99.3\% of improvement
to the use of the Compton reflection model. 

\begin{figure}
\epsfig{file=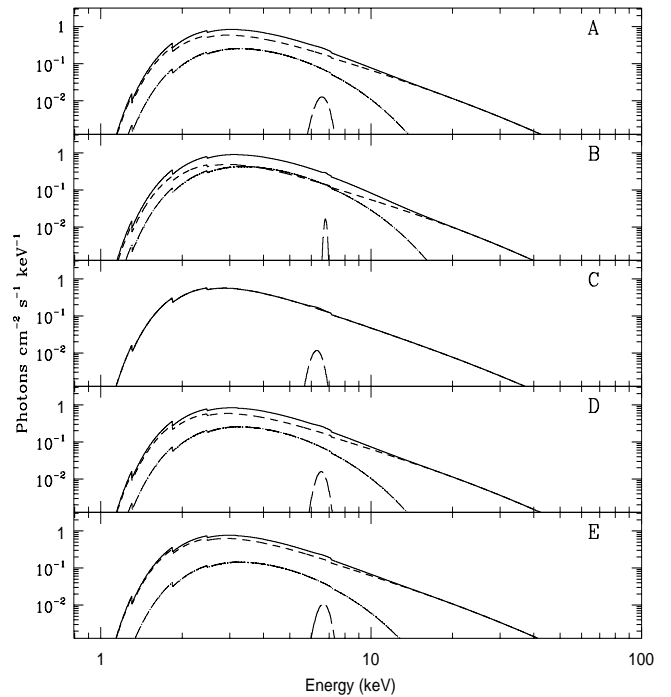,width=9cm,height=10cm}
\caption[]{Best fit models for the broad band
BeppoSAX spectra in the five sections shown in figure 1.
The model include an absorbed (N$_{h} \sim 5.6\times10^{22}$)
power law with a high energy cut-off and a 
Compton reflection component from a disk (dashed curve), 
a multi-temperature disk blackbody (dot-dash),
and an Iron line (long dash). The continuous line shows the
sum of the various components.}
\label{Models}
\end{figure}

\begin{figure}
\epsfig{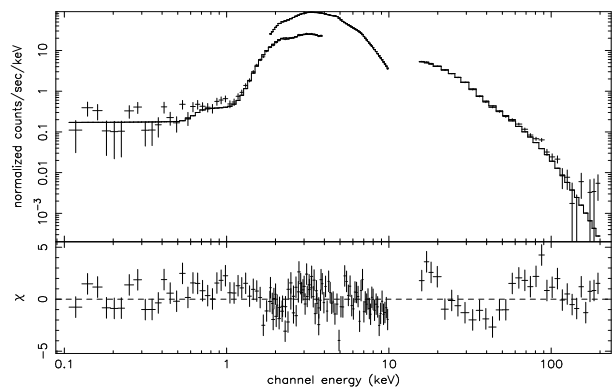}
\caption[]{Data and best fit model for the broad band
BeppoSAX spectrum of interval E. Count-rates from all the three instruments
(LECS, MECS and PDS) are shown here together with their residuals
with respect to the model (in units of individual contributions to the
$\chi^{2}$)  
}
\label{Spectrum}
\end{figure}

In Fig.~\ref{Models} we show the best-fit models to the data of 
the five selected time intervals, showing both the individual components 
and the total spectrum. For the sake of clarity we did not show real data
in this figure, whose purpose is to show the evolution of the individual
spectral components. As an example of the statistical quality of our
data we show the actual data for interval E - for 
which we also have data from the LECS - in Fig.~\ref{Spectrum},
together with the best fit model and residuals (we note that this
is one of the cases with the worse $\chi^{2}$). 

In Tables 1 and 2 we report the spectral fit parameters
(all errors are given at a 90\% confidence level, unless
otherwise specified).\footnote{Even if the
value for the reduced $\chi^{2}$ is rather high in some intervals, namely
A and E, we were not able to get better fits, either changing parameters
or changing the model. We therefore think that these values are due
to intrinsic rapid variation of the continuum on timescales shorter
than those used here ($\sim$500-3000~s), and/or 
to residual systematic uncertainties in the deconvolution matrices of the
instruments, because of the high intensity of this source. 
We stress, however, that we
just used the statistical counting errors in the fit, without
adding any systematic error.}
A word a caution is necessary here for what concerns the physical
meaning of the inner disk radius and temperature.
First of all, general relativity effects (e.g. \cite{zhang97}) 
are not included in the adopted model ({\it diskbb} in {\it XSPEC}).
For instance, as discussed by Sobczak et~al. (1999a, 1999b), 
the relativistic corrections imply a radius of the innermost stable orbit 
smaller than what derived from this model. 
Also, the {\it diskbb} model may be not appropriate to describe the 
source energy spectrum in some conditions (see also Merloni et al. (1999)
for a critical discussion of this effect), resulting in an unrealistic
value for the inner disk radius.
Therefore, the values of inner radius given in Table~2 have to be 
regarded with extreme caution.
(We note, however, that these effects do not substantially affect
our observational evidence of the lack of the thermal component during
interval C.) 

In Table~2 and Fig.~\ref{Models}
it is immediately evident that the soft thermal component (the disk) 
significantly contribute to the spectrum in intervals A, B, D and E,
while in the dip (interval C) it is not required by the data.   
During the outburst (interval B) it shows the largest intensity and 
temperature, while it appears basically constant in A, D and E.
The power law component also shows significant variations.
In particular, the power law photon index varies only slightly
along the event, but for the interval C where it shows a increase of $\sim$0.3-0.4.
Correspondingly, the high energy cut-off shows a significant increase,
passing from $\sim$50-60~keV to $\sim$100~keV.
Unfortunately, the parameter uncertainties for the Fe line and the 
reflection component are too large to search for 
possible variability.

\begin{table*}[hbt]
\setlength{\tabcolsep}{1.5pc}
\caption{Best-fit parameters for the continuum. The model includes 
an interstellar absorption (fixed at 5.6$\times10^{22}$ cm$^{-2}$,
a power law with high energy cut-off, a Compton reflection component, 
an emission iron line and (except for interval C)
a blackbody emission from a multi-temperature accretion disk.
Errors are given at a 90\% confidence level for one parameter.}
\begin{tabular*}{\textwidth}{@{}c@{\extracolsep{\fill}}cccccc}
\hline
\hline
Time     & Power law              & Cut-off               & Inner disk           & Inner disk             
& Compton                  & Reduced       \\
Interval & index                  & Energy (keV)          & radius$^{(*)}$ (km)  & Temp. (keV)            
& Reflection$^{(**)}$      & $\chi^{2}$ (d.o.f.) \\ 
\hline
A        & 2.36$^{+0.05}_{-0.04}$ & 49.6$^{+5.3}_{-3.9}$  & 25.7$^{+1.6}_{-1.7}$ & 1.53$^{+0.06}_{-0.05}$ 
& $<$ 0.36               & 1.99 (89)\\ 
B        & 2.33$^{+0.12}_{-0.08}$ & 42.8$^{+10.5}_{-6.4}$ & 27.0$^{+1.7}_{-1.3}$ & 1.69$^{+0.05}_{-0.05}$ 
& $<$ 0.81               & 1.34 (89)\\ 
C        & 2.68$^{+0.03}_{-0.03}$ & 94.3$^{+29.6}_{-19.1}$ &   -                 & -                      
& 0.47$^{+0.30}_{-0.26}$ & 1.02 (92)\\ 
D        & 2.44$^{+0.07}_{-0.06}$ & 58.8$^{+9.5}_{-6.6}$  & 26.4$^{+2.5}_{-2.1}$ & 1.51$^{+0.07}_{-0.07}$ 
& 0.41$^{+0.35}_{-0.26}$ & 1.51 (89)\\ 
E        & 2.53$^{+0.04}_{-0.06}$ & 70.2$^{+9.1}_{-8.8}$  & 19.5$^{+3.9}_{-1.7}$ & 1.52$^{+0.08}_{-0.15}$ 
& 0.47$^{+0.24}_{-0.27}$ & 1.68 (89)\\ 
\hline
\end{tabular*}
$^{(*)}$ Derived from the XSPEC model {\it diskbb}, assuming a distance of 12~kpc 
and a inclination angle to the disk of 70$^{\circ}$\\
$^{(**)}$ Dimensionless parameter {\it rel$\_$refl} of the XSPEC {\it pexrav} model, corresponding to
the solid angle subtended by the reflecting matter to the illuminating source in units of 2$\pi$
\end{table*}

\begin{table*} [hbt]
\setlength{\tabcolsep}{1.5pc}
\caption{Best-fit parameters for the iron line. The model is the same as in Table 1.}
\begin{tabular*}{\textwidth}{@{}c@{\extracolsep{\fill}}cccccc}
\hline
\hline
Time     & E$_{line}$      &  $\sigma$              & Line flux              & Equivalent width  \\
Interval & keV             & keV                    & 10$^{-2}$ ph/cm$^2$/s  & eV  \\
\hline
A & 6.57$^{+0.14}_{-0.14}$ & 0.35$^{+0.19}_{-0.19}$ & 1.25$^{+0.67}_{-0.50}$ &        45 \\ 
B & 6.72$^{+0.28}_{-0.29}$ & $<$ 0.48               & 0.57$^{+0.64}_{-0.45}$ &        19 \\ 
C & 6.29$^{+0.24}_{-0.25}$ & 0.3~(fix)              & 0.97$^{+0.41}_{-0.41}$ &        55 \\ 
D & 6.53$^{+0.11}_{-0.20}$ & $<$ 0.56               & 1.24$^{+0.88}_{-0.63}$ &        45 \\ 
E & 6.65$^{+0.13}_{-0.25}$ & 0.35$^{+0.41}_{-0.20}$ & 1.00$^{+1.28}_{-0.41}$ &        44 \\ 
\hline
\end{tabular*}
\end{table*}

\begin{table*}[hbt]
\setlength{\tabcolsep}{1.5pc}
\caption{
Best-fit values used for the correlation presented in Fig.~\ref{pl_index_qpo} and 
\ref{flux_qpo}:
QPO centroid frequency, power law flux (2-10~keV),
flux (2-10~keV) from the multi-temperature disk blackbody and
power law flux (15-50~keV) 
}  
\begin{tabular*}{\textwidth}{@{}c@{\extracolsep{\fill}}cccccc}
\hline
\hline
Time     & QPO Centroid        & PL flux$^{b}$ & DBB flux$^{b}$   & PL flux$^{b}$ \\
Interval & Frequency$^{a}$~(Hz)& (2-10 keV)    & (2-10 keV)       & (15-50 keV)   \\ 
\hline
A        & 5.23$\pm0.11$        & 2.71          & 1.05             & 0.70          \\ 
B        & -                    & 2.25          & 1.82             & 0.59          \\ 
C        & 4.12$\pm0.18$        & 2.36          & -                & 0.47          \\ 
D        & 5.16$\pm0.13$        & 2.62          & 1.04             & 0.69          \\ 
E        & 4.58$\pm0.08$        & 2.74          & 0.59             & 0.65          \\ 
\hline
\end{tabular*}
$^{a}$ The centroid value was obtained from a gaussian+polynomial fit to the QPO peak
and the local underlying continuum. The confidence level of the associated error is 3$\sigma$.\\
$^{b}$ Unabsorbed flux in units of 10$^{-8}$ erg cm$^{-2}$ s$^{-1}$, corresponding to
a luminosity of $\sim1.6\times10^{38}$ erg s$^{-1}$ 
\end{table*}

\section{Timing analysis}

\begin{figure}
\epsfig{file=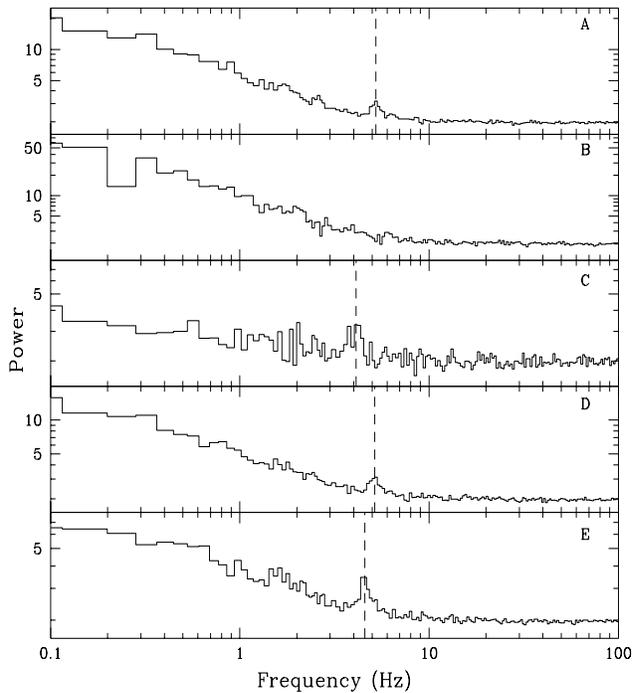,width=9cm,height=10cm}
\caption[]{Power Spectral Density obtained by the MECS data for the
five intervals given in Fig.\ref{LC_flare}.
Frequency bins are spaced according to a geometric series.
 The vertical dashed lines indicate
the centroid of the QPO frequency at 
5.23~Hz (A), 4.12~Hz (C), 5.16~Hz (D) and 4.58~Hz (E).
}
\label{Psd}
\end{figure}

We used the MECS (1.8--10~keV) 
data to study the temporal behaviour of the
source in the five intervals. The MECS light curves 
where produced with a 3 ms bin size. Using the Xronos package (v. 4.02) we 
divided the light curves in time segments $\sim$12~s long
and calculated the power density spectrum for each of them.
The spectra reported in Fig.~\ref{Psd} are then
obtained averaging a minimum of $\sim$40 power density spectra
(up to 280 for the longest interval).

A peaked feature is detected at a frequency between 4 and 5~Hz, varying
both in frequency and intensity over the 5 intervals. 
We interpret such a feature as a quasi-periodic oscillation (QPO)
of the `intermediate type' (\cite{morgan97,markwardt99,muno99}).
The QPO is detected at 5.23~Hz in interval A, then is 
not detected during the outburst (interval B), it appears again
at 4.12~Hz during the dip (interval C),
and moves to 5.16~Hz and 4.58~Hz during intervals D and E, respectively.
It is interesting to note how the feature gets sharper in interval E, 
while it seems broader in interval C.
In this interval the QPO peak could actually be the
convolution of a variable peak QPO, as suggested by observations
of similar source states by RXTE (e.g., \cite{markwardt99,muno99}), where
the QPO frequency is observed to drift in time on very short timescales.
Also, it is worth noting the possible existence of a small peak in interval A at 
a frequency $\sim$2.6~Hz, possibly a sub-harmonic of the QPO as 
observed in other black hole candidates (see van der Klis 1995 for a
review). 

It is worth noticing that the shape of the low frequency continuum 
is significantly different in interval C (the dip, during
which the disk component is not detected in the energy spectrum), 
where it is much flatter than in the other intervals, particularly with 
respect to B (the outburst) where it appears the highest.

\section{Discussion}

The spectral analysis of the BeppoSAX X--ray data 
confirmed that the radio/X flare event that we observed
is consistent with the scenario, proposed by Belloni et al. (1997a),
in which X-ray outbursts occur when the inner part of the accretion disk
is disrupted, and plasma is expelled out of the disk, thus generating
the radio peak observed at 15~GHz.  
The global behaviour of the spectral parameters is also
consistent with what found by Trudolyubov et al. (1998) on similar 
source states.

\begin{figure}
\epsfig{file=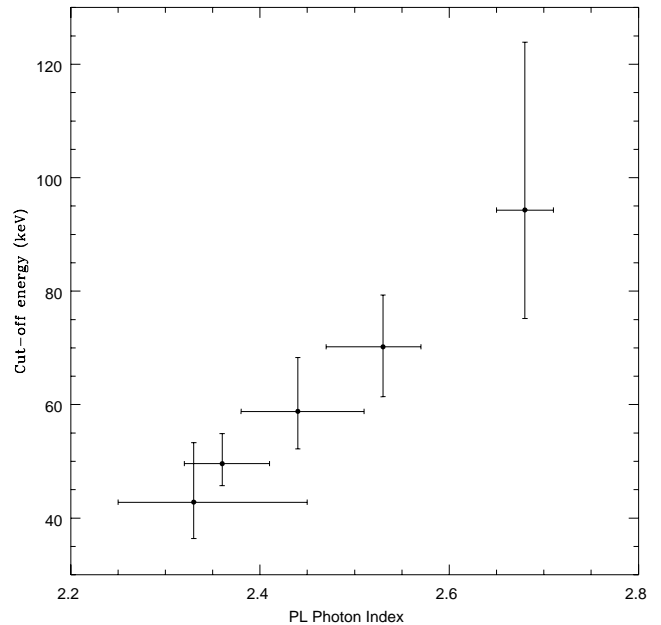,width=9cm}
\caption[]{Correlation between the power law photon index and
the high energy cut-off
}
\label{pl_cut}
\end{figure}

We studied possible correlations 
between spectral and timing parameters. 
First, we find that the photon index of the power law
component appears to be positively correlated with the cut--off energy
(Fig.~\ref{pl_cut}). A word of caution is necessary here, 
as the two parameters are strongly correlated in 
the fit procedure, and this is exactly the kind of correlation that could
arise if the power law index were wrongly estimated for some 
unknown reason. 
On the other hand, as we will show below the
power law photon index appears well correlated to temporal parameters too,
which have been derived from a completely independent analysis. Even if this
cannot be considered as a proof, it is nevertheless  a suggestion in 
favour of the correctness of the measurement of the 
power law indices, and therefore of the reality of the correlation.
We have also searched the literature for a confirmation of this
effect. 
Heindl et al. (1997) and Muno et al. (1999)
use data from the HEXTE experiment onboard RossiXTE.
In table 2 of the former paper we find only two useful fits, and they
are consistent with our correlation.
In table 3 of Muno et al. (1999) we find
six spectral fits in which our correlation is not found.
However, we note that the value of the cut-off energy in the latter
analysis is always larger ($\sim$100~keV, and not far from 
the upper energy bound of the data set) than what we find,
while Heindl et al. (1997) obtain values similar to ours (60 and 88~keV).

The timing analysis of our X--ray data 
shows the presence of a variable QPO around 5~Hz.
This frequency is classified as `intermediate' in the phenomenology
of this source, 
and the centroid frequency is usually observed to vary 
between 1.5 and 15~Hz.
Trudolyubov et al. (1998) and Muno et al. (1999), 
analysing data from RossiXTE with the source in a variety of spectral
and intensity states,
found positive
correlations of the intermediate QPO centroid frequency with the low energy 
flux from the power law component and from the thermal disk emission.
Also, Markwardt et~al. (1999) found a correlation between the 
frequency of the intermediate QPO and the parameters of the disk component. 
 
\begin{figure}
\epsfig{file=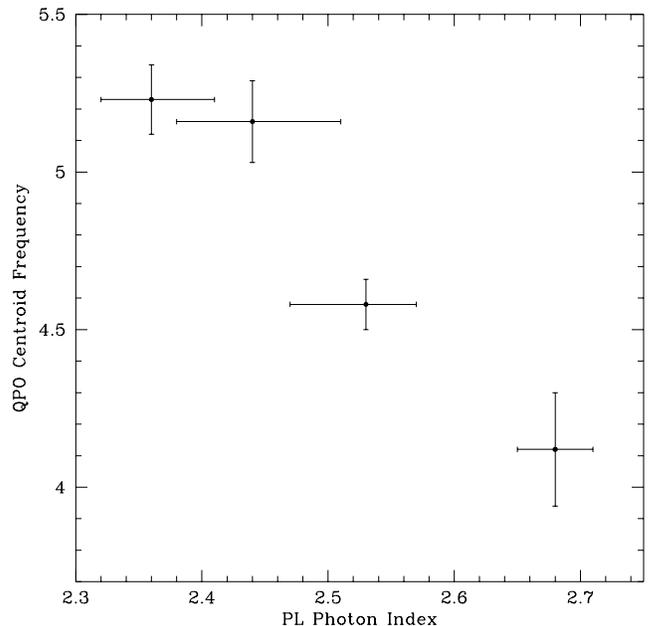,width=9cm}
\caption[]{
Correlation between the power law photon index and the centroid frequency 
of the QPO
}
\label{pl_index_qpo}
\end{figure}

\begin{figure}
\epsfig{file=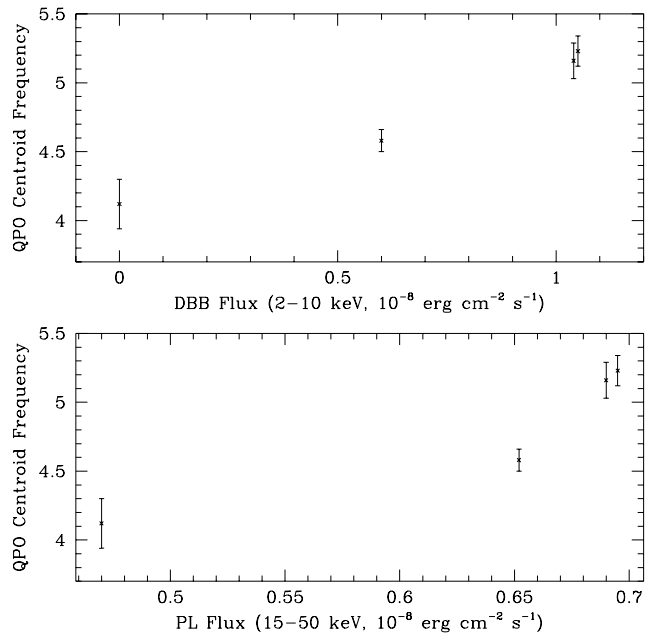,width=9cm}
\caption[]{
{\it Top Panel:} 
Correlation between the 2-10~keV flux from the multi-temperature disk blackbody 
and the centroid frequency of the QPO\\
{\it Bottom Panel:} 
Correlation between the flux in the 15-50 keV range from the power law and
the centroid frequency of the QPO
}
\label{flux_qpo}
\end{figure}

In our data set a negative correlation exists between the QPO 
centroid frequency and the power law photon index (Fig.~\ref{pl_index_qpo}).
Given the correlation between the power law index and the cut-off
energy it is also possible that the real correlation is
between the QPO frequency and the cut-off energy. 
In principle, it could be related to the already known QPO-flux correlation, 
but we only find a correlation between the QPO centroid frequency and the flux 
in the 2--10~keV from the disk blackbody and in the 15--50~keV 
from the power law component 
(Fig.~\ref{flux_qpo}). On the contrary, there is not clear correlation with 
the low energy (2--10~keV) flux from the power law component, even if our small
data set does not allow to definitely rule it out. 
This 
may be explained if the basic correlation is with the photon index, 
and the power law is pivoting at low energies; actually, the variation 
in the 2--10~keV flux is significantly lower than in the 15--50~keV flux.

We note that our results on the correlations of the QPO centroid
frequency with the disk blackbody flux and the high energy power law
flux are in agreement with what found by Muno et al. (1999) using RXTE
data, while the anti-correlation with the power law index is  
opposite to what can be inferred by using the different correlations
presented by these authors.
This might be a chance occurrence due to our small data set or
an intrinsic difference due to the fact that the analysis of the RXTE
data includes a large number of observations and source states, whereas
our data refer to a single flare event and are integrated over longer
time segments.

\section{Conclusions}

Our observations clearly indicate that the GRS 1915+105 
system is subject to overall instabilities affecting
both the disk and its surroundings.
We studied an isolated radio flare
(possibly caused by an expanding cloud of energized electrons
emitting at large distance from the central object)
associated with an isolated
X--ray outbursting activity
followed by a shallow dip.
It is remarkable that the relative timing of the radio and  
X--ray flare in this episode appears possibly different from what 
seen in some observations presented in the literature 
(see Sect.~2.1).
The X--ray outbursting episode (lasting in our data $\sim600$~s)
appears to be chaotic with rapid fluctuations that most
likely coincide with the energization of an external corona
and of the radio emitting electrons.
The ultimate cause of this instability is unknown.
Several suggestions have been proposed, including
thermal/viscous disk instabilities (\cite{taam97}) and/or
turbulent magnetic field dissipation in the inner
part of the accretion disk (\cite{tavani96}).
The different states that we observe for the power law component
might be caused by a quasi-thermal and moderately thick
Comptonizing corona, or by an optically thin and magnetically energized
particle population radiating by synchrotron and/or inverse Compton.
Our observations do not favor one or the other mechanism.

Based on the observational results obtained from both the
spectral and timing analysis of the BeppoSAX data, 
we try to phenomenologically interpret the observed source 
behaviour in the framework of the model proposed by 
Belloni et al. (1997a). 
The relevant observational results are:\\ 
(a) a correlation exists between the QPO centroid frequency and the spectral 
    parameters of both the disk and the power law component;\\
(b) the disk emission is maximum during the X--ray outburst (interval B); \\
(c) high amplitude, fast oscillations are detected in X--rays during the outburst;\\
(d) the QPO is absent during the outburst;\\
(e) the low-frequency noise is maximum during the flare and it is minimum
during the dip;\\
(f) the disk emission disappears during the dip (interval C); \\ 
(g) the lowest QPO frequency is observed during the dip;\\
(h) the cut-off energy jumps to its highest value during the dip;\\
(i) the radio emission rise together with the X--ray outburst, and 
    peaks during the dip;\\
(l) the thermal disk emission is restored after the dip (intervals D and E).

Our point (a) indicates that the three observed source characteristics (disk
emission, power law emission, QPO) either originate from the same site, 
or in different sites causally related with each other.
(For example, it has been 
proposed (\cite{chakrabarti95,molteni96,kazanas97,titarchuk98}) 
that a 1--10~Hz QPO might originate from radial oscillations 
at a boundary between the 
optically thick accretion disk and an extended atmosphere 
of hot electrons, responsible for the power law emission through comptonization
of soft disk photons.)

At the time of the outburst the disk undergoes violent instabilites (point c)
causing plasma emission out of the disk. 
The thermal emission from the disk is therefore reinforced 
by the plasma emission (point b). When the plasma is ejected
into the extended atmosphere it goes through the boundary region 
where the QPO
might take place, disturbing the dynamical equilibrium that generates it,
and causing its temporary disappearance (point d). 
After the plasma jet ejection the internal disk no longer exists/emits
(point f). The ejected plasma has thus caused an increase of the average
energy of the coronal electron population, 
detected both as an increase of the cut-off energy 
(point h), and as a smooth increase of the radio synchrotron 
emission (point i) that reaches its peak when the ejection of plasma
out of the disk ends. 
The dynamical equilibrium at the boundary region, 
temporarily perturbed by the 
plasma ejection event, is now being restored.
The quasi-periodic oscillation therefore starts again, but it is
at a lower frequency (point g), possibly due to a
a larger distance from the central object, as also suggested
by Muno et al. (1999) who find a positive correlation between 
the innermost disk temperature and the QPO frequency.
Then, at the time of intervals D and E,
a new dynamical equilibrium is reached when the inner disk is filled again
and is detectable at X--rays (point l).
It is also interesting that the low-frequency noise appears significantly 
lower when the inner disk is absent and maximum when the emission from the disk 
is at its highest value (point e), demonstrating that this timing feature is
associated with the inner disk. 
  
Therefore, the results obtained from our analysis of an isolated flare
are consistent with a `complete cycle' of an accretion disk perturbation,
causing the temporary disappearance of the inner portion of the disk,
within a scenario in which the accretion disk and the Comptonizing 
electrons population are geometrically or physically related. 
The interaction between the two would be therefore responsible for 
the quasi-periodic oscillation of intermediate ($\sim$5~Hz) frequency.
The even more complex behaviour often observed in this source, could
then be a combination (with possible modifications and complications
and on different timescales) of cycles similar to the one observed here.

\begin{acknowledgements}

We wish to thank R. Hjellming for useful discussions about radio jets 
and the anonymous referee for comments that helped to improve the 
clarity of the paper.
The BeppoSAX Science Operation Center and Science Data Center 
assisted the authors during the observations and data reduction. 
EC, MF and GM acknowledge financial support by ASI and MURST.
\end{acknowledgements}

\end{document}